\documentclass[onecolumn,aps,superscriptaddress,letterpaper,amsmath]{revtex4}
\usepackage{graphicx}
\usepackage{pdfpages}
\usepackage[utf8]{inputenc}
\usepackage{hyperref}

\begin{document}

\title{Constructing coarse-grained skyrmion potentials from experimental data with Iterative Boltzmann Inversion}
\author{Yuqing Ge}
\thanks{Y.G. and J.R. contributed equally to this letter.}
\affiliation{Institut für Physik, Johannes Gutenberg-Universität, Staudinger Weg 9, D-55099 Mainz, Germany.}
\affiliation{Department of Physics, Chalmers University of Technology, Göteborg 41296, Sweden}
\author{Jan Rothörl}
\thanks{Y.G. and J.R. contributed equally to this letter.}
\author{Maarten A. Brems}
\author{Nico Kerber}
\author{Raphael Gruber}
\author{Takaaki Dohi}
\author{Mathias Kläui}
\thanks{Authors to whom correspondence should be addressed: klaeui@uni-mainz.de, virnau@uni-mainz.de}
\author{Peter Virnau}
\thanks{Authors to whom correspondence should be addressed: klaeui@uni-mainz.de, virnau@uni-mainz.de}
\affiliation{Institut für Physik, Johannes Gutenberg-Universität, Staudinger Weg 9, D-55099 Mainz, Germany.}

\begin{abstract}
In an effort to understand skyrmion behavior on a coarse-grained level, skyrmions are often described as 2D quasi particles evolving according to the Thiele equation. Interaction potentials are the key missing parameters for predictive modeling of experiments. We apply the Iterative Boltzmann Inversion technique commonly used in soft matter simulations to construct potentials for skyrmion-skyrmion and skyrmion-magnetic material boundary interactions from a single experimental measurement without any prior assumptions of the potential form. We find that the two interactions are purely repulsive and can be described  by an exponential function for experimentally relevant skyrmions. This captures the physics on experimental time and length scales that are of interest for most skyrmion applications and typically inaccessible to atomistic or micromagnetic simulations.
\end{abstract}

\maketitle

Magnetic skyrmions are two-dimensional whirls of magnetization stabilized by their non-trivial topology. They have been observed in bulk materials and magnetic thin films at temperatures ranging from a few Kelvin to far above room temperature \cite{Fert2017, Finocchio2016, EverschorSitte2018, Jiang2015, Nagaosa2013, Jiang2017, Lindner2020, Huang2020, Woo2016}. Skyrmions can be stabilized even without the application of an external field \cite{Lemesch2018, Zheng2017}, can be manipulated efficiently with spin torques \cite{Jonietz2010, Yu2012, Woo2016} and exhibit thermally activated dynamics \cite{Huang2020, Song2021, Zazvorka2019}. This turns them into attractive candidates for applications in novel devices relying on skyrmion-based logic \cite{Zhang2015a, Zazvorka2019}, Brownian computing \cite{Nozaki2019, Brems2021} and racetrack memory \cite{Fert2013, EverschorSitte2018, Zhang2015b, Tomasello2014, Schaeffer2020}. High density lattice states, in which skyrmions strongly interact with each other, have recently moved into the focus of attention \cite{Muehlbauer2009, Nakajima2017, Huang2020, Zazvorka2020, Ognev2020}. In contrast to colloids \cite{Zahn1999, Zahn2000, Keim2007}, sizes and density of skyrmions can be adjusted on the fly which makes them ideal model systems to study phase behaviour in two dimensions \cite{Kapfer2015} as demonstrated by the demonstration of hexatic and solid phases in Refs. \cite{Huang2020, Zazvorka2020}.

To describe theoretically skyrmion ensembles that for the experimentally relevant systems spans length scales over microns to hundreds of microns, conventional micromagnetic approaches are not applicable due to the prohibitive computational cost. Therefore coarse-grained particle-based descriptions have been introduced \cite{Lin2013, Brown2018, Brown2019, Schaeffer2019, Foster2019, Brearton2020, Reichhardt2015, Reichhardt2016, Reichhardt2021}, where skyrmions are represented by repulsive soft disks whose dynamics are governed by the Thiele equation \cite{Thiele1973}. While this approach is successful in describing individual skyrmion dynamics \cite{Zhao2020, Zazvorka2020}, modelling of skyrmion ensembles or of skyrmion movement in tight confinements \cite{Brems2021, Schaeffer2020} requires knowledge of the interaction potentials. So far, interaction potentials between skyrmions are mostly rationalized by theoretical considerations. 
If one considers skyrmions as quasi-particles, then the description in the simplest approximation requires interaction potentials between a skyrmion and a sample edge as well as between a skyrmion and another skyrmion. While possible other interactions can be envisaged, one would start with these two simple descriptions and then compare to experiments to see if these suffice to describe the dynamics of multiple skyrmions in confined geometries.
So far, the skyrmion interaction potentials have been analyzed based on a micromagnetic energy functional. By determining the spin-structure of a skyrmion, interaction potentials between two such structures and between a skyrmion and the magnetic material boundary have been approximated by a potential which is exponentially decreasing with distance \cite{Lin2013, Foster2019, Brearton2020}. Concrete parameters for coarse-grained particle-based potentials at the nanometer scale are obtained by fitting micromagnetic simulation data to this functional form \cite{Lin2013,Schaeffer2019}.
Ref.~\cite{Zazvorka2020} followed a different approach by directly matching structural information from experimental high density states with corresponding particle-based simulations. Instead of an exponential function, a repulsive power-law was used to enable comparisons with previous simulations of phase transitions in Ref. \cite{Kapfer2015}. This approach, however, entails the severe limitation that the interaction potentials are purely repulsive and therefore unable to capture possible attractive parts in the interaction as have recently been suggested by micromagnetic simulations \cite{Rosza2016}.
Methods like Reverse Monte Carlo \cite{Lyubartsev1995} or Iterative Boltzmann Inversion (IBI) \cite{Reith2003, Moore2014}, which are rooted in computational soft matter physics, do not require such assumptions and are inherently better suited for the analysis of the full interaction potentials. In these approaches, potentials are adjusted in successive simulations to match a target pair-correlation function, and employed to construct coarse-grained computational models from more detailed atomistic descriptions. Even though the possibility of experimental reference functions are often discussed in this context \cite{Milano2005}, they have so far been rarely employed.

Building upon Ref. \cite{Zazvorka2020}, we here propose a method with which we can derive system-specific coarse-grained potentials for skyrmion-skyrmion and skyrmion-boundary interactions from the analysis of a single measurement without applying assumptions about the shape of the interaction potentials using the Iterative Boltzmann Inversion method. By determining potentials this way, we find that structural properties of a skyrmion lattice with $\mu m$-size skyrmions are consistent with soft repulsive potentials.

\begin{figure}[htbp]
\includegraphics[width = \textwidth]{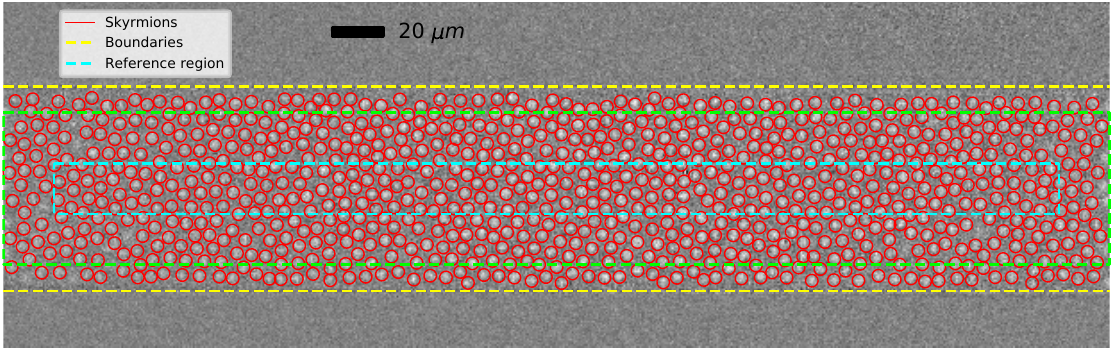}
\caption{MOKE image of an experimental skyrmion conformation. 
The field of view covers part of a long rectangular stripe of magnetized material along the horizontal direction with a width of $80.68\ \mu m$. The two parallel yellow dashed lines denote the location of the boundaries of the magnetized area. Skyrmions are displayed as bright bubbles on grey background and marked by red circles as detected using the Trackpy module \cite{Trackpy, Crocker1996} for python. The diameter of the circles is about $5\ \mu m$ and smaller than the actual interaction range. The region enclosed by the cyan dashed line denotes the area from which skyrmions are taken as reference particles in the determination of the pair-correlation function. In the region enclosed by the green dashed line, skyrmions are still counted as neighbors for skyrmions from inner part for the calculation of $g(r)$.
The distance between cyan and green dashed lines is $20\ \mu m$ to keep $g(r)$ in figure~\ref{fig:GofR} properly normalized in the range of interest.
\label{fig:ExperimentalPicture}}
\end{figure}

\begin{figure*}[t]
\includegraphics[width = \textwidth]{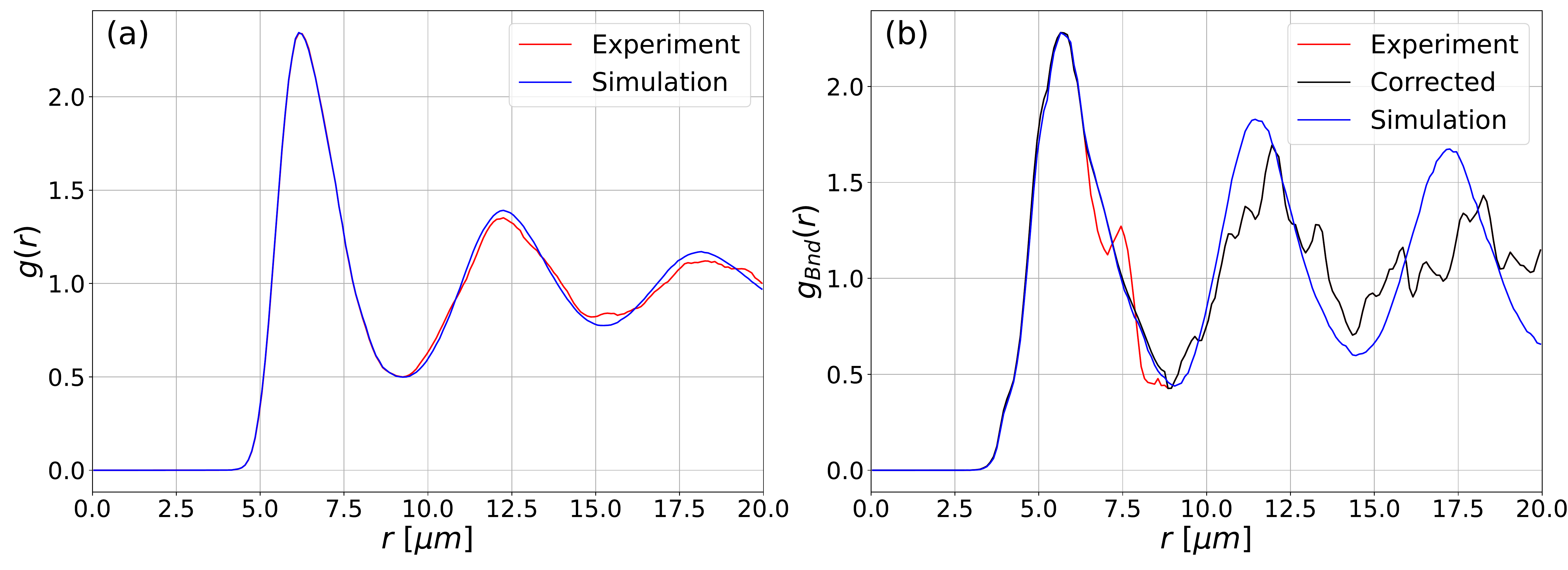}
\caption{Plots of experimental and simulated radial distribution functions for skyrmion-skyrmion \textbf{(a)} and skyrmion-boundary interactions \textbf{(b)}. The simulated distribution function is generated by a MD simulation after performing 500 update steps as described in the main text. \textbf{(b)} additionally contains a corrected experimental function which is used to determine the simulation potential. The bin width $\Delta r$ of of all functions is 0.1 $\mu m$.
\label{fig:GofR}}
\end{figure*}

Skyrmion interactions were analyzed in a $\text{Ta}(5)/\text{Co}_{20}\text{Fe}_{60}\text{B}_{20}(0.9)/\text{Ta}(0.08)/\text{MgO}(2)/\text{Ta}(5)$ (layer thickness in $n$m in parentheses) multilayer stack which had already been used in observation of skyrmion lattices in confined geometries \cite{Song2021}. The sample exhibits ferromagnetic properties with interfacial Dzyaloshinskii–Moriya interaction (DMI) and perpendicular magnetic anisotropy (PMA). The out-of-plane (OOP) component of magnetization was imaged using polar magneto-optical Kerr effect (MOKE) microscopy.
Rectangles of magnetic material were patterned on the sample with electron beam lithography (EBL) and Argon ion etching (figure~\ref{fig:ExperimentalPicture}). The width of the rectangle was 81 $\mu$m as estimated from the background image. On the two short sides, the boundaries were far enough outside of the analyzed region to have negligible edge effects.
The temperature of the sample ($341.5K$) was controlled by a Peltier element and tuned by the current source. This temperature allows for creating a number of well visible and detectable skyrmions with a density that can be controlled by field-cycling. Skyrmions were nucleated after an in-plane magnetic field pulse of $30~mT$ which saturated the magnetization in the plane under a fixed OOP field of $70~\mu T$.
Particle tracking was performed with the python module TrackPy \cite{Trackpy} (see Supplementary Material for more information on the tracking procedure), and a trajectory video was taken for 20 minutes at a frame-rate of 16 frames per seconds. As skyrmion density still decreased in the first five minutes and stabilizes afterwards, only the remaining 15 minutes of the measurement were used for analysis.\\
From this data we computed the average of the one-dimensional radial distribution function for each frame (figure~\ref{fig:GofR}a):
\begin{equation}
    g(r) = \frac{1}{\rho N_\text{i}} \frac{1}{2 \pi r \Delta r} \sum_{i \neq j} \left[ \theta(r_{ij} - r) - \theta(r_{ij} - r - \Delta r) \right].
    \label{eq:GofR}
\end{equation}
Here, $\rho$ is the skyrmion area density calculated from the region inside the green dashed lines in figure~\ref{fig:ExperimentalPicture}. $N_\text{i}$ is the number of skyrmions enclosed by the cyan lines and $\Delta r$ the bin width of the radial distribution function. $\theta$ refers to the Heaviside-function and the sum over $i \neq j$ indicates summation over all pairs of particles $i$ within the cyan and particles $j$ within the green dashed line in figure~\ref{fig:ExperimentalPicture}. This function describes the normalized probability distribution of distances between skyrmions with the first peak corresponding to nearest neighbors. The boundary distribution function (figure~\ref{fig:GofR}b) was calculated in a similar manner via
\begin{equation}
    g_\text{Bnd}(r) = \frac{1}{\rho} \frac{1}{L \Delta r} \sum_{i} \left[ \theta(r_{i} - r) - \theta(r_{i} - r - \Delta r) \right],
    \label{eq:GofRBnd}
\end{equation}
where $r$ is the distance in the direction perpendicular to the boundary and $L$ is the length of the boundary in the recorded video. The yellow line in figure~\ref{fig:ExperimentalPicture} shows the position of the boundary determined from the position of average brightness between darker parts outside and brighter parts inside the rectangle.
The boundary distribution function for the two boundaries exhibited a noticeable shift of $0.6\ \mu m$, which can be explained by drifting of the sample compared to the background image used to determine the positions of the boundary. Because the exact amplitude of the drift cannot be determined directly using the Kerr videos, we shift both functions by $0.3\ \mu m$ so that the first maximum is located at the same position assuming that the two boundaries interact identically with skyrmions. The boundary distribution function determined this way is further corrected by calculating the running average over eight neighbors in both directions for the region between the first peak and the first minimum of $g_\text{Bnd}$. This is done because the small peak at around $7.5~\mu m$ in the experimental function, likely caused by pinning-induced quenched disorder, does not contribute to the real shape of the skyrmion-boundary interaction of this system. Note that each skyrmion only contributes once per frame to the skyrmion-wall correlation function, while for skyrmion-skyrmion interactions pairs are considered, which is the main reason for the difference in statistics. The applied correction is shown in figure~\ref{fig:GofR}b and used for all simulations. 

To derive interaction potentials for the given experimental setup we performed particle-based Brownian dynamics simulations  \cite{Brown2018, Brown2019, Lin2013} according to the Thiele equation \cite{Thiele1973}
\begin{equation}
    -\gamma \vec{v} - G \hat{z} \times \vec{v} = \vec{F}_\text{SkSk} + \vec{F}_\text{SkBnd} + \vec{F}_\text{Random}
\end{equation}
with temperature $k_B T = 1$, damping $\gamma = 1$ and a Magnus force amplitude $G = 0.25$ as already used in a previous work \cite{Song2021}. $\vec{F}_\text{Random}$ are forces arising from thermal white noise satisfying the fluctuation-dissipation relation. The width of the simulated system and the skyrmion density are set according to experimentally determined values. Periodic boundary conditions are used in x-direction.
\\
This particle-based model of skyrmions comes with some limitations as it averages over effects like skyrmion deformations, internal excitation modes, size polydispersity, creation and annihilation of skyrmions, and also long-range and multi-particle interactions. Deformations and excitation modes, however, mostly occur under applied currents \cite{Litzius2017} or fluctuating magnetic fields \cite{Onose2012, Beg2017} and are therefore less relevant to the analysis of static properties of a confined lattice. This model is considered a trade-off between the fidelity of treating the skyrmion structures, the complexity of the model and the robustness of the results with limited data typically available experimentally. Therefore, it is used most robustly for analyzing static properties like lattice formation and ordering or dynamics under small applied forces for systems with constant numbers of skyrmions. Its application is very versatile as it can easily be scaled to small skyrmion numbers in confinement \cite{Song2021} or the analysis of large lattices \cite{Zazvorka2020}. We note that the model currently assumes circular spin structures and in our experiment, skyrmions are monodisperse in size and show no eccentricity in any preferred direction thus making the model applicable. For skyrmions that are deformed by external in-plane fields as realized previously \cite{newKerber2021}, the model would indeed require adaptation. 

Interaction potentials for skyrmion-skyrmion and skyrmion-boundary interactions are determined using the Iterative Boltzmann Inversion method \cite{Reith2003}. The first simulation is undertaken with the potential of mean force
\begin{equation}
    V_0(r) = - k_B T \ln g(r)
\end{equation}
for the respective distribution functions $g(r)$ and $g_\text{Bnd}(r)$ where $r$ denotes the distance between two skyrmions or between a skyrmion and the boundary, respectively. Potentials are calculated this way up to the first minimum of the distribution function and set to zero for larger distances. Using the potentials $V_{\text{SkSk},0}$ and $V_{\text{SkBnd},0}$, a Molecular Dynamics (MD) simulation is performed to obtain the distribution functions $g_0(r)$ and $g_{0,\text{Bnd}}(r)$. Those are used to simultaneously update the interaction potentials by
\begin{equation}
    V_{i+1} (r) = V_i (r) + \alpha k_B T \ln \frac{g_i(r)}{g(r)}
\end{equation}
where $\alpha \in (0,1]$ is a parameter determining the speed of updating the potential which is chosen to be 0.2 in this work. The two updated potentials are then used for the next simulation. This process is repeated until the potential converges, and in total 500 updating iterations each including MD simulations were performed.
The IBI method has two important advantages over similar methods used in previous publications \cite{Zazvorka2020}. First, no assumptions on the shape of the potential must be provided for the simulations. Instead the shape is obtained directly from the experimental data. Second, IBI allows for determining skyrmion-skyrmion and skyrmion-boundary potentials simultaneously by updating the two in every step.\\
Results for the radial distribution functions are displayed in figure~\ref{fig:GofR}. The method is able to reproduce both distribution functions very closely in the observed region and also to a good extent beyond this region at least for the skyrmion-skyrmion interaction. The different shape of the skyrmion-boundary interaction beyond the fitted region is most likely a consequence of pinning-induced quenched disorder which is not considered in the simulation.

\begin{figure*}[htb]
\includegraphics[width = 0.5\textwidth]{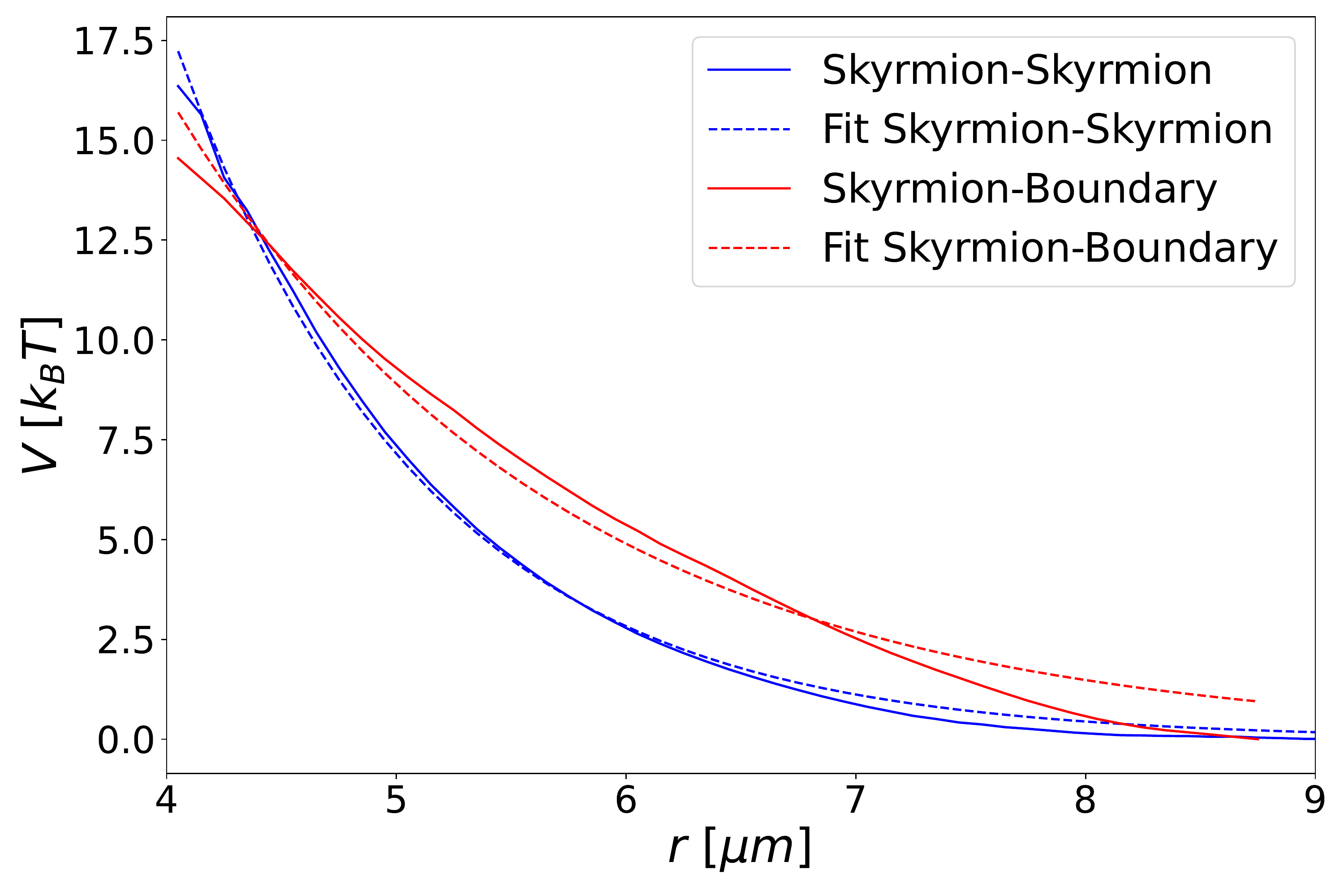}
\caption{Skyrmion-skyrmion and skyrmion-boundary potentials determined by Iterative Boltzmann Inversion and fits of the potentials to an exponential function $V(r) = a \exp(-r / b)$ in the region from 4.0 $\mu m$  up to the first minimum of the respective distribution function. Fit values are $a_\text{SkSk} = 735.1~k_B T$, $b_\text{SkSk} = 1.079~\mu m$, $a_\text{SkBnd} = 176.7~k_B T$, and $b_\text{SkBnd} = 1.673~\mu m$.
\label{fig:Potentials}}
\end{figure*}

Potentials resulting from this method are displayed in figure~\ref{fig:Potentials}. The two potentials we determined for this experimental system are purely repulsive and can be approximated by exponential functions. While agreement with the generic functional form predicted from theoretical considerations and micromagnetic simulations for skyrmions on the nanometer scale \cite{Lin2013, Brearton2020} is good for our extracted skyrmion-skyrmion interactions, somewhat larger deviations can be observed for skyrmion-boundary interactions. It is worth noting that IBI is in principle well-suited for capturing attractive interactions~\cite{Reith2003}. Particle-based simulations may even be adapted on-the-fly to scenarios in which skyrmion potentials change from attractive to repulsive as described in \cite{Du2018} or in which the skyrmion sizes change as a result of applied fields as long as radial distribution functions for the extreme positions are accessible.
\\
Finally, we would like to test the capability of our potentials to reproduce independent static properties of the experimental system. To this end, we turn to structural quantifiers which describe phases and ordering of 2D quasi particles. We start with the local orientational order parameter \cite{Kapfer2015} $\psi_6 (j) = 1/n_j \sum_{k = 1}^{n_j} e^{6i \theta_{jk}}$, where $n_j$ represents nearest neighbors of skyrmion $j$ as determined by Voronoi tessellation, and $\theta_{jk}$ is the angle between the line connecting skyrmions $j$ and $k$ and the $x$-axis \cite{Kapfer2015}. Considering skyrmions in the cyan reference region defined in figure~\ref{fig:ExperimentalPicture}, the absolute value of the local order parameter $\langle |\psi_6| \rangle$ as averaged over all skyrmions and snapshots is $0.486 \pm 0.008$ for the experimental system and $0.496 \pm 0.007$ for the simulated system. Similarly favorable agreement can be obtained by the percentage of particles with six neighbors $P_6$. For the experiment we obtain an average $P_6 = (59.2 \pm 0.8)\%$ and $P_6 = (62.0 \pm 1.1)\%$ for the simulation.

\begin{figure*}[htb]
\includegraphics[width = 0.75\textwidth]{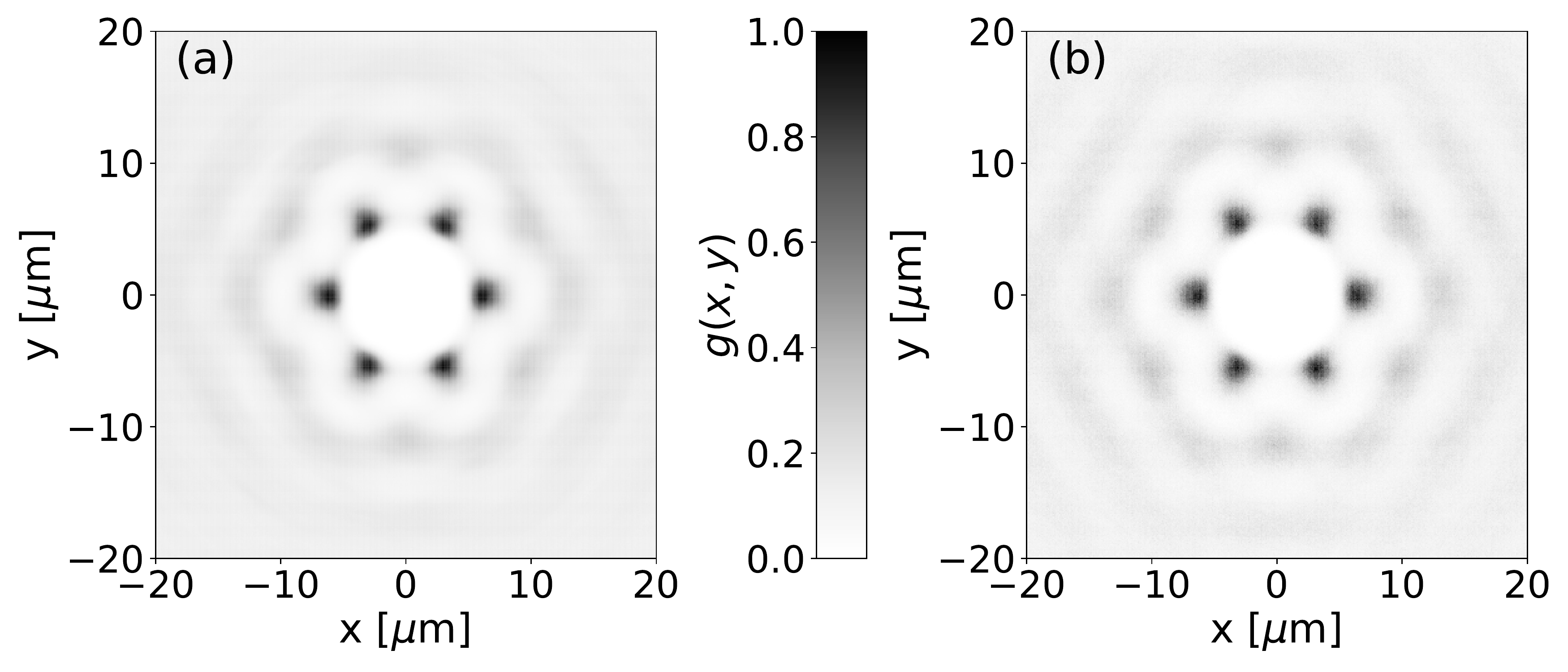}
\caption{(a) Normalized 2-dimensional radial distribution function $g(x,y)$ of the experimental system. (b) Normalized 2-dimensional radial distribution function $g(x,y)$ of the simulated system.}
\label{fig:2DGofR}
\end{figure*}
The ordering of skyrmions beyond the first shell can visualized by the two-dimensional radial distribution function $g(x,y)$ shown in figure~\ref{fig:2DGofR} which is calculated with the same constraints as in 1D. The six peaks around the center show the preferred local sixfold coordination of the system while the nearly isotropic second circle implies absence of long-range orientational order.
We conclude that the system is in the liquid phase \cite{Zazvorka2020,Ognev2020} as already indicated in figure~\ref{fig:ExperimentalPicture}. Agreement between experiment and simulation with respect to structural quantifiers is actually quite remarkable (considering that pinning effects have been neglected in the simulations) and confirms the capability and accuracy of our approach. It also shows that pinning effects are reduced in skyrmion lattices due to the skyrmion-skyrmion interaction.

In conclusion, we derive for the first time skyrmion-skyrmion and skyrmion-boundary potentials for particle-based simulation models directly from experimental data and without any assumptions on the shape of the potential. We find that the two interactions can be described by repulsive potentials, thus bridging the gap in time and length scales between micrometer-sized experimental skyrmions and theoretical predictions possible only at the nanoscale. Thereby our work captures the physics on experimental time and length scales where fully atomistic or micromagnetic simulations are typically not feasible, but which are relevant for devices based on skyrmions. Our result allows for more accurate simulations of micrometer-scale skyrmion systems without being compelled to use potentials determined for inherently different systems. Our procedure is well-suited to construct coarse-grained simulation models for concrete experimental setups and are a first step towards the quantitative exploration of novel skyrmion devices with computer simulations.
The method can also be applied to a broad range of experimental systems. These include not only the multilayer stacks leading to skyrmion lattices as studied here but also much smaller skyrmions that can be studied in bulk systems hosting skyrmions \cite{Huang2020} making this a universally applicable method as long as pair correlation functions are accessible.
The latter can also be accessed for nano-scale skyrmions, which cannot be easily imaged by microscopy. They can however be characterized by neutron diffraction \cite{Muehlbauer2009, Nakajima2017}. During such experiments the scattering pattern can be easily converted to the pair distribution function by a Fourier transform. Thus, the approach of extracting skyrmion-skyrmion interaction potentials from a pair distribution function has the potential to be applied to both, nano-scale \cite{Huang2020} and micro-scale skyrmions \cite{Zazvorka2020}. As a consequence, one can use it to experimentally validate interaction models between skyrmions including those with partially attractive interactions that so far have only been predicted by micro-magnetic simulations.

A comprehensive and quantitative description of experiments will also require a description of the effects of non-flat energy landscapes \cite{Gruber2022}, which could be integrated into our approach as well as dynamical properties. Moreover, analyzing skyrmion systems at different densities and comparing results between these systems could provide further insights in the properties of skyrmion interactions that might go beyond the two-body description explored in this work.

\appendix
\subsection*{Acknowledgements}
P.V. acknowledges fruitful discussions with Martin Hanke-Bourgeois. We are grateful to the Deutsche Forschungsgemeinschaft (DFG, German Research Foundation) for funding this research: Project number 233630050-TRR 146 and Project number 403502522-SPP 2137 Skyrmionics. The authors furthermore acknowledge funding from TopDyn, SFB TRR 173 Spin+X (project A01 \#268565370), and from the Horizon 2020 framework program of the European commission under grant No. 856538 (ERC-SyG 3D MAGIC).

\subsection*{Author contributions}
Y.G. and J.R. contributed equally to this letter. P.V. and M.K. devised the study. N.K. fabricated the sample. Y.G., N.K. and R.G. conducted the experiments. Y.G. and J.R. evaluated the experimental data. J.R., Y.G. and M.B. performed simulations. T.D., M.K. and P.V. supervised the study. All authors contributed on preparing the manuscript. 

\section{Supporting Information}
\subsection{Locations of particles and confinements }
\begin{figure}[htbp]
	\includegraphics[width = 0.5\linewidth]{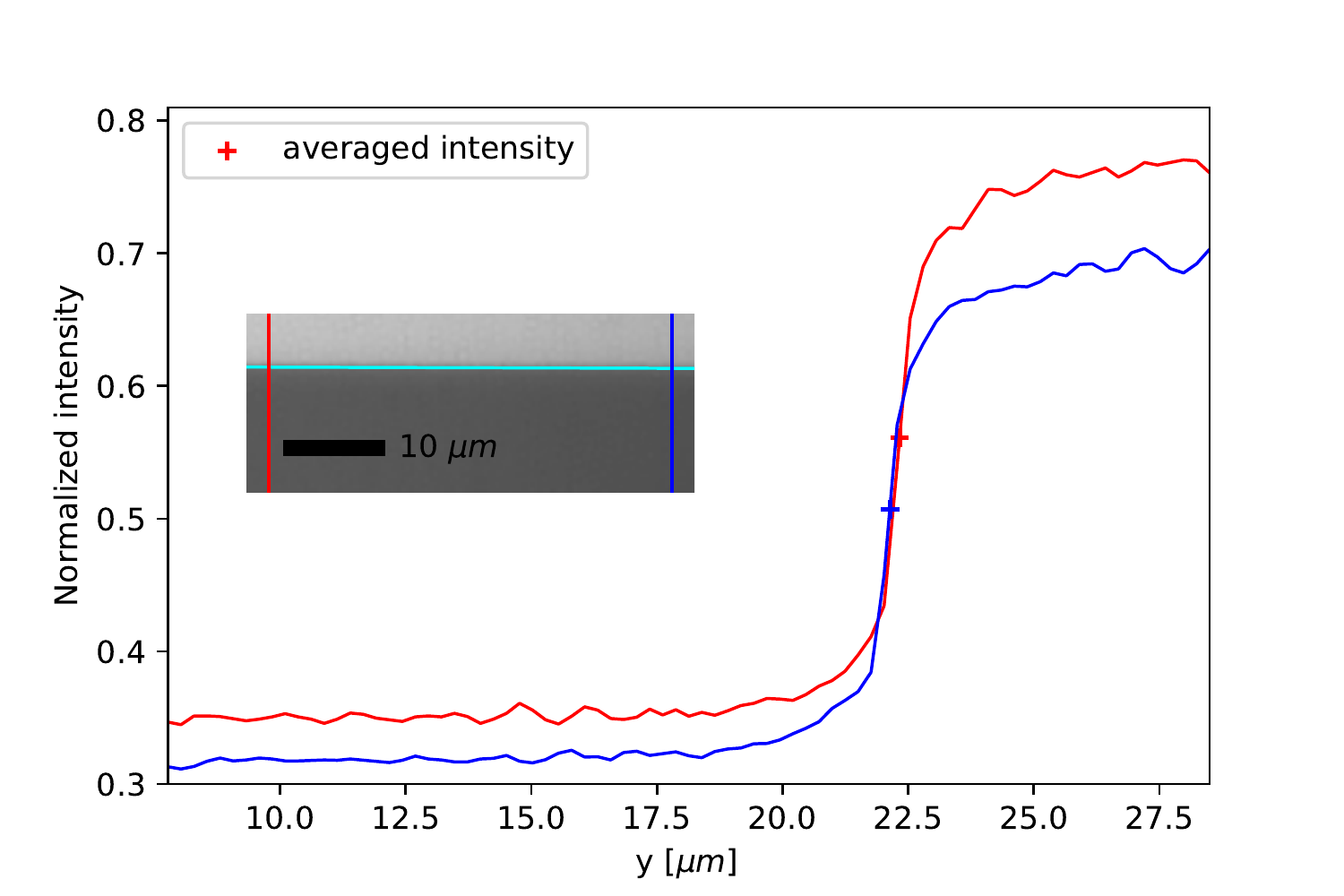}
	\caption{Location of a confining boundary. The inset is a picture taken by a Kerr microscope containing the boundary between the magnetized material exhibiting higher intensity and non-magnetized material of lower intensity. The cyan horizontal line is the located boundary. Exemplary, intensities along the two vertical lines in the inset are plotted as a function of the y-coordinate in the figure with corresponding colors. 
		We calculated the average within the two regions and found coinciding positions which are indicated by the crosses. The difference between the two curves in y-direction is a result of uneven illumination. This intensity gradient does not affect location of the boundary since the average value was calculated along each line separately.
		\label{fig:Walllocations}}
\end{figure}

Locations of skyrmions were obtained from Kerr images with the Python module Trackpy \cite{Trackpy} which locates particles in grey-scale images. First, we used Fast Fourier transformation to determine the average distance between skyrmions, which was then employed as diameter of the masks to locate skyrmions using the Trackpy module. The mask diameter was generally slightly larger than skyrmion bubbles such that they are fully covered. The frames were pre-processed to remove short-wavelength noises by Gaussian blurring, and long-wavelength noises such as the uneven illumination. In the presented configuration, we used a mask diameter of $9\ px$, which is equivalent to $5.84\ \mu m$, a Gaussian blurring kernel of $1.5\ px$, and a separation $4\ px$. The locations were then refined by least-square fitting with a Gaussian function.\\
In the calculation of distances between skyrmions and confining boundaries, the locations of boundaries were determined from a Kerr image of the sample taken at the beginning of the measurement. The magnetized foreground has a higher intensity than the background. Thus we compute where the intensity equals the average of the fore- and background as in figure~\ref{fig:Walllocations}, then fit the points with a polynomial of first order. However, this average from the Kerr microscope does not indicate the boundary of the material directly, and the sample drifted thermally during the measurement. This could lead to a disagreement between $g_{Bnd}(r)$ of two boundaries. Considering the symmetry of potentials from both confining boundaries, we shifted the boundary locations globally by $0.3~\mu m$ so that the first peak of $g_{Bnd}(r)$ of both boundaries coincide. 

\vfill
\newpage
\bibliographystyle{apsrev4-1}
\bibliography{references}

\end{document}